# Chalcogenide Metasurfaces Enabling Ultra-Wideband Detectors from Visible to Mid-infrared

Shutao Zhang[1,2,3,#], Shu An[1,#], Mingjin Dai[4], Qing Yang Steve Wu[1], Nur Qalishah Adanan[2], Jun Zhang[1], Yan Liu[1], Henry Yit Loong Lee[1], Nancy Lai Mun Wong[1], Ady Suwardi[1,5], Jun Ding[3], Robert Edward Simpson[2,6,*], Qi Jie Wang[4,*], Joel K. W. Yang[2,*] and Zhaogang Dong[1,2,3,*]

[1]Institute of Materials Research and Engineering (IMRE), Agency for Science, Technology and Research (A*STAR), 2 Fusionopolis Way, Innovis #08-03, Singapore 138634, Republic of Singapore

[2]Singapore University of Technology and Design (SUTD), 8 Somapah Road, 487372, Singapore

[3]Department of Materials Science and Engineering, National University of Singapore, 9 Engineering Drive 1, Singapore 117575

[4]School of Electrical and Electronic Engineering, Nanyang Technological University, Singapore 639798, Singapore

[5]Department of Electronic Engineering, The Chinese University of Hong Kong, Sha Tin, New Territories, Hong Kong SAR 999077, China

[6]University of Birmingham, Edgbaston, B15 2TT, UK

[#]These authors equally contribute to this work.

*Correspondence and requests for materials should be addressed to J.K.W.Y. (email: joel_yang@sutd.edu.sg), Q.J.W. (email: qjwang@ntu.edu.sg), R.E.S. (email: r.e.simpson.1@bham.ac.uk) and Z.D. (email: dongz@imre.a-star.edu.sg).

**ABSTRACT**

Thermoelectric materials can be designed to support optical resonances across multiple spectral ranges to enable ultra-wide band photodetection. For instance, antimony telluride ($Sb_2Te_3$) chalcogenide exhibits interband plasmonic resonances in the visible range and Mie resonances in the mid-infrared (mid-IR) range, while simultaneously possessing large thermoelectric Seebeck coefficients. In this paper, we designed and fabricated $Sb_2Te_3$ metasurface devices to achieve resonant absorption for enabling photodetectors operating across an ultra-wideband spectrum, from visible to mid-IR. Furthermore, relying on asymmetric $Sb_2Te_3$ metasurface, we demonstrated the thermoelectric photodetectors with polarization-selectivity. This work provides a potential platform towards the portable ultrawide band spectrometers at room temperature, for environmental sensing applications.

KEYWORDS: Ultrawide band, $Sb_2Te_3$ metasurface, thermoelectric effect, photodetector

**Introduction**

Metasurfaces, consisting of sub-wavelength nanostructures arranged on a two-dimensional manner, offer capabilities for manipulating light-matter interactions. It therefore enables a powerful platform to achieve compact optical devices for various applications, such as ultra-thin optical lens,[1] nanostructured color pixels,[2] communication,[3] fluorescence enhancements,[4] optical nonlinearity,[5] anti-counterfeiting,[4b, 6] energy harvesting,[7] and miniaturized optical detectors.[8] For example, due to its resonant interaction with light, miniaturized optical detectors that are integrated with semiconductor metasurfaces are able to detect the multi-dimensional characteristics of light, such as wavelength,[8d] polarization,[9] and angle.[10] Nevertheless, these detection mechanisms are usually constrained by their bandgap to a specific working spectral range, either to visible, near-IR or mid-IR range.

On the other hand, photodetection based on thermoelectric effect has the intrinsic advantage of wide working wavelength range. For example, aluminum plasmonic metasurfaces fabricated on top of a commercial photo-thermoelectric (PTE) detector enhances the light absorption and photodetector sensitivity for both visible and near-IR spectra.[11] Other thermoelectric material platforms, such as 2D $MoS_2$, $WSe_2$ and $PdSe_2$, have been explored,[12] for the integration with either nanoantenna[13] or waveguide[14], with reports of relatively low light absorptance.[15] For instance, absorptance of graphene can be as low as 2.3%.[16] An alternative approach uses the lateral $p$-$n$ heterojunction based on $Bi_2Te_2Se$-$Sb_2Te_3$, with both the flat film design[12a] and grating structures with guided mode resonances,[17] mostly working in the visible spectrum. Some of us have recently demonstrated PTE detectors integrated with an optical cavity for mid-IR photodetection.[18] However, the full potential of thermoelectric material flatform for ultra-wideband light detection, especially from visible to mid-IR, has not been explored.

In this paper, we utilize the optical and thermoelectric properties of a chalcogenide material, antimony telluride ($Sb_2Te_3$) to design ultra-broadband metasurface detectors operating at room temperature. We leverage on its interband plasmonic resonance, showing examples for the visible spectrum and Mie resonance in the mid-IR spectrum, via exploring their respective resonant absorption capabilities. The photodetection mechanism involves photo-absorption that leads to local temperature rise and a measurable voltage across the material due to its thermoelectric properties. By leveraging the Seebeck coefficient of approximately 178 µV/K and the dielectric function supporting $Sb_2Te_3$ Mie resonances[19] and plasmonic resonances, we designed ultra-wideband metasurfaces with resonant wavelengths and achieved a maximum absorptance of approximately 97% at ~532 nm. Additionally, the $Sb_2Te_3$ metasurface was employed to create polarization-selective ultrawide band thermoelectric photodetectors. This research contributes the integration of thermoelectric and optical functionalities through nanofabrication techniques. It provides precise control of device absorptance over wavelength and polarization, aiming to improve photodetection applications.

## Results

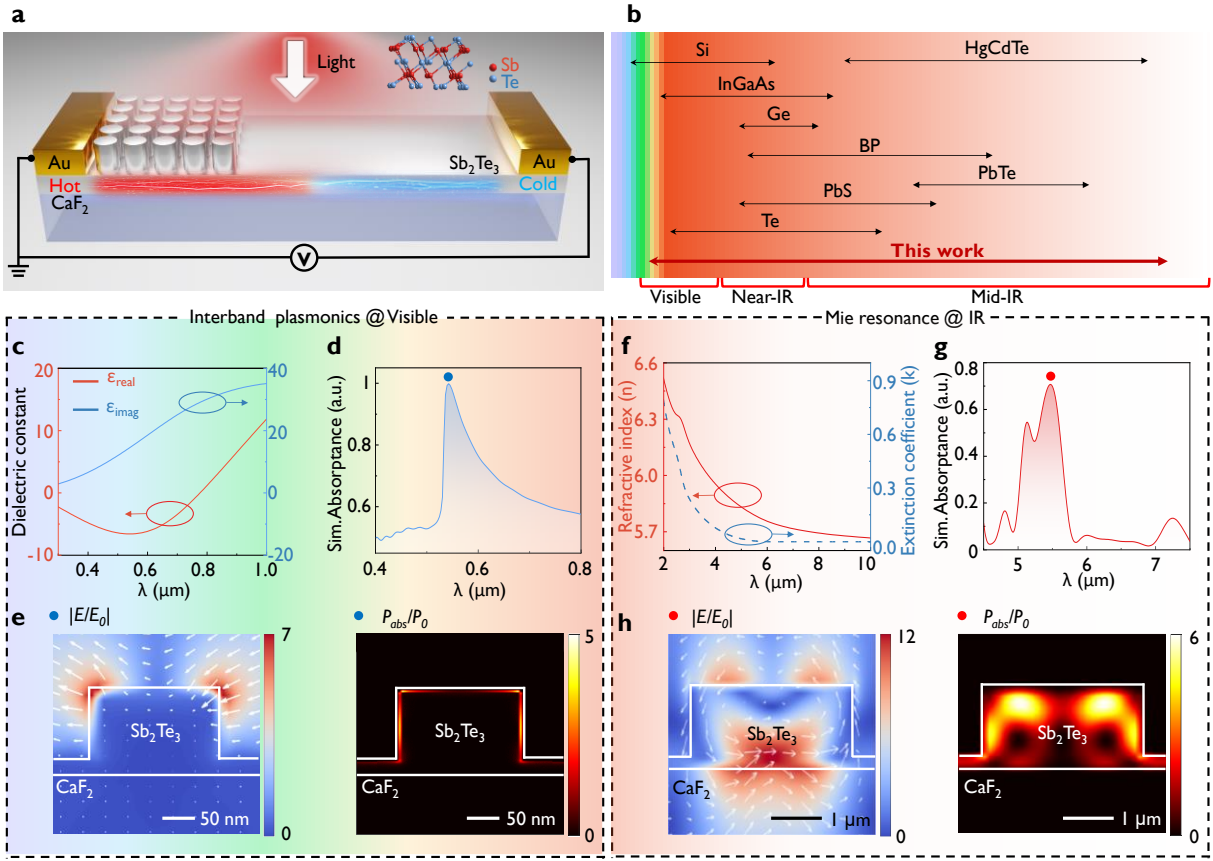

**Figure 1. Sb$_2$Te$_3$ chalcogenide metasurfaces for enabling ultra-wideband photodetectors due to the resonant absorption from visible to mid-IR spectrum.** (a) Schematic of the Sb$_2$Te$_3$ metasurface photodetectors consisting of nanoposts being etched into the Sb$_2$Te$_3$ film close to the left electrode. The enhanced absorptance at the metasurface region will lead to a temperature difference upon light absorption and a thermoelectric-induced measurable voltage readout across the electrodes. The insert shown the crystal structure of rhombohedral-phase Sb$_2$Te$_3$. (b) A brief summary and benchmarking on the operational wavelength range of common material platforms for photodetectors. (c) The dielectric function of Sb$_2$Te$_3$ film in the visible region, where the real part is negative in this band. (d) Visible absorptance spectrum of the Sb$_2$Te$_3$ nanostructures (diameter: 200 nm, height: 150 nm and pitch: 532 nm). (f) The refractive index (*n*) and extinction

coefficient ($k$) of $Sb_2Te_3$ in mid-IR range. (g) Simulated mid-IR absorptance spectrum of the $Sb_2Te_3$ nanostructures (diameter: 2.0 μm, height: 900 nm and pitch: 3.0 μm). (e, h) Simulated electric field distributions normalized to incident electric field ($|E/E_0|$) distribution and absorption intensity distribution ($|P_{abs}/P_0|$) across a cross-section of the visible (blue point) and mid-IR (red point) $Sb_2Te_3$ nanodisk.

Figure 1a presents the schematic of our designed thermoelectric photodetector that leverages on the combined thermoelectric and optical properties of $Sb_2Te_3$ nanostructures. The illustration shows the crystal structure of $Sb_2Te_3$, and the material characterization results are shown in Figure S1. The detector consists of a $Sb_2Te_3$ strip on $CaF_2$ substrate, being connected to Au electrodes at both ends for electrical voltage readout. To generate a temperature gradient upon light illumination, $Sb_2Te_3$ metasurface is patterned only at one end of the strip. This metasurface enhances light absorptance and thus the metasurface will be hotter than the unpatterned $Sb_2Te_3$ side, inducing a voltage difference. To achieve wavelength-selective detection, $Sb_2Te_3$ nanostructures with varying structural parameters can be designed, enabling resonant light absorption to specific wavelengths across a broad spectrum. Figure 1b presents the working wavelength range of our designed $Sb_2Te_3$ metasurface detector, spanning a larger range than bandgap limited devices of other photodetector material platforms.[20] Noted that the responsivity of thermoelectric photodetectors are notoriously low, but remains constant throughout.[21]

First, we discuss the plasmonic resonance characteristics of $Sb_2Te_3$ in the visible spectrum, where Figure 1c presents the dielectric constant. The corresponding refractive index ($n$) and extinction coefficient ($k$) of $Sb_2Te_3$ in visible range is depicted in Figure S2. At wavelengths ranging from 300 to 760 nm, the real part of the dielectric constant is negative, indicating a plasmonic range due to interband transitions.[22] For instance, Figure 1d presents the simulated absorption spectrum of $Sb_2Te_3$ nanodisk arrays, which shows a plasmonic resonance with a peak at 532 nm. At this wavelength, the electric field near the material surface is enhanced by ~7, as illustrated in Figure 1e (left panel). This enhanced field further promotes the interaction between light and $Sb_2Te_3$, thereby increasing the absorption effect, as shown in Figure 1e (right panel). Characteristic of plasmonic-like resonances, the absorption is constrained to the surface of the nanostructure. Note that the dielectric constant of our grown $Sb_2Te_3$ thin film is different from the one as reported in the literature,[22b] and such discrepancy arises from the film deposition conditions.

Next, we discuss the Mie resonance characteristics of $Sb_2Te_3$ at mid-IR range due to its high refractive index, which enables the mid-IR field localization due to the high Q-factor resonances. For instance, Figure 1f presents the refractive index (*n*) and extinction coefficient (*k*) in the mid-IR range, where it has a high refractive index of 5.7-6.5 at the wavelength range of 2-10 μm. Finite-difference time-domain (FDTD) simulations were carried out for mid-IR $Sb_2Te_3$ nanodisk arrays (diameter: 2.0 μm, height: 900 nm and pitch: 3.0 μm), where the simulated absorptance spectrum is shown in Figure 1g. The absorptance peak at 5.8 μm indicates resonant absorption, demonstrating its capacity for mid-IR resonant interactions, where the corresponding electrical field amplitude ($|E/E_0|$) and absorption distribution are shown in Figure 1h. The high electric and magnetic fields within the $Sb_2Te_3$ disk indicate strong intensity localization, exhibiting characteristics of Mie resonance. Additionally, we need the 150-nm-thick $Sb_2Te_3$ base layer for electrical conduction. At the same time, this $Sb_2Te_3$ base layer also introduces the Fabry-Pérot (FP) resonance. Therefore, the interaction between these two resonances leads to hybrid Mie-FP resonances. To better study the optical response mechanism of the metasurface, multipolar decomposition analysis was conducted, as shown in Figure S3, where magnetic dipole (*MD*) is the dominant component, followed by electrical dipole (*ED)* contributions at 5.8 μm.

Here, we also would like to mention that In the near-IR region, $Sb_2Te_3$ acts as a lossy dielectric material without effective structural resonances, leading to inferior detector performance compared to its capabilities in the visible and mid-IR spectra, as indicated by the experimental measurements of the complex refractive index *n* and *k* as shown in Figure S4. At the same time, due to the intrinsic material absorption, $Sb_2Te_3$ should be still able to work as PTE detector at near-IR wavelength region.

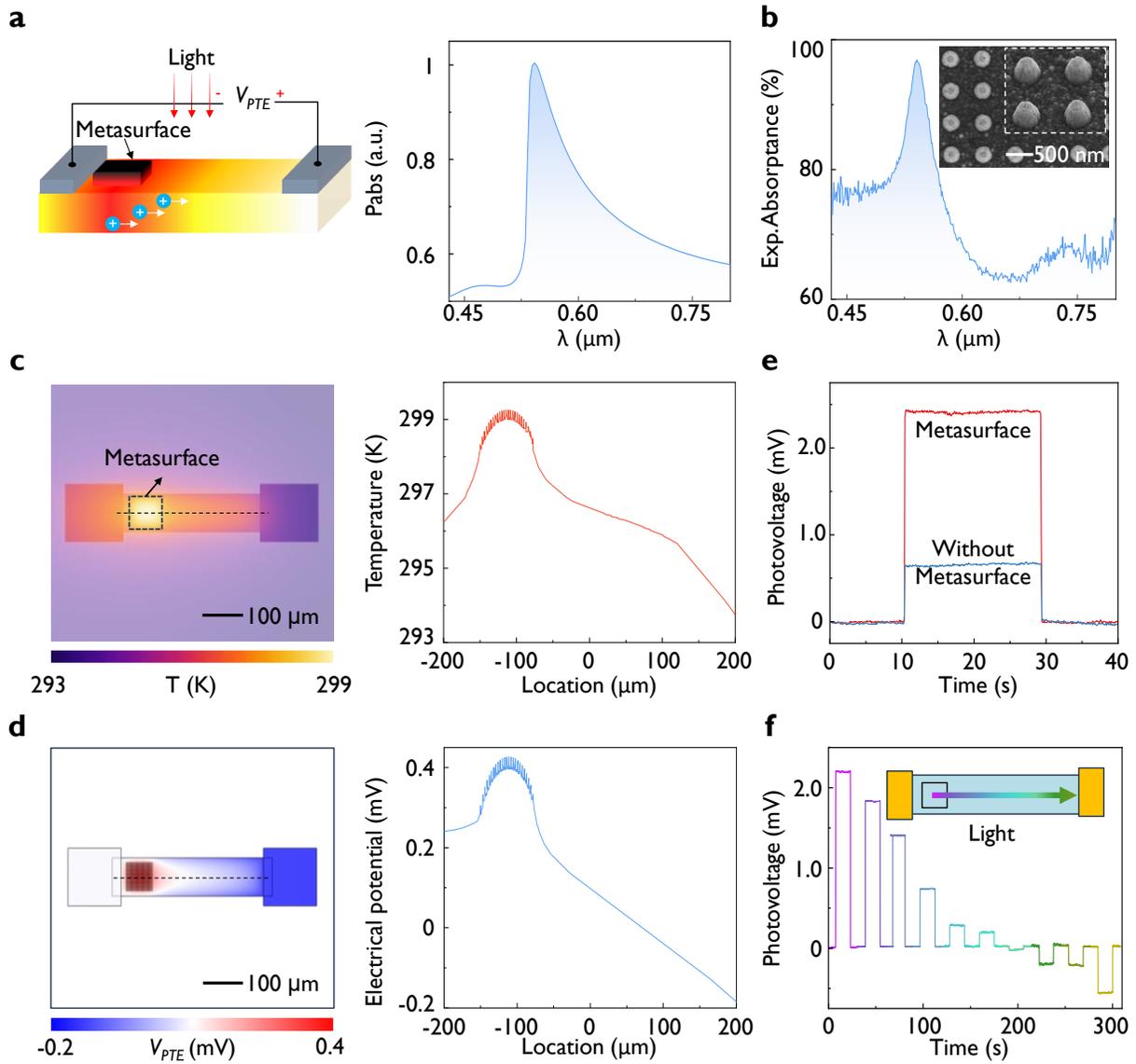

**Figure 2. Chalcogenide metasurface photodetectors based on interband plasmonics of Sb$_2$Te$_3$ at visible wavelength.** (a) Schematic illustration of the photo-thermoelectric (PTE) conversion process in the metasurface. Simulated absorption power density with an incident power of 0.1 mW for the visible light metasurface (Right panel). (b) Measured absorptance spectrum of the designed visible Sb$_2$Te$_3$ photodetector (disk diameter 200 nm, height 150 nm, pitch 532 nm, base Sb$_2$Te$_3$ film thickness of 150 nm). The visible Sb$_2$Te$_3$ nanostructures and the base layer is shown insert. (c) Simulated temperature gradient with an input heating power of 0.1 mW and corresponding

temperature map for the thermoelectric detector with the $Sb_2Te_3$ metasurface. (d) Simulated resulting electric potential and corresponding potential map. Metasurface occupies an area of 75×75 μm² and consists of disks of diameter 200 nm, height 150 nm, pitch 532 nm. (e) Measured photovoltages from the $Sb_2Te_3$ detector when the 532 nm laser illuminates the metasurface and flat substrate regions respectively. The source meter unit is grounded at the electrode near the metasurface. (f) Measured photovoltage when the focused 532 nm laser spot is scanning from metasurface to flat region. The incident optical power is 0.1 mW.

Based on the PTE effect, the photoresponse can be divided into two distinct processes: photothermal effect and thermoelectric effect, as illustrated in Figure 2a. When the active material, such as $Sb_2Te_3$, absorbs light, a temperature difference is generated between the two terminals of the device. The majority carriers (holes in the *p*-type material $Sb_2Te_3$) are driven by the Seebeck effect from the hot end to the cold end, resulting in a potential difference between the two terminals of the channel. For photothermal conversion, we enhance the local heating through the hybrid resonances of the metasurface. Figure 2a (right panel) shows the simulated absorption power density of the $Sb_2Te_3$ metasurface in the visible region, respectively, with the strongest absorption observed at 532 nm. Within the visible wavelength range, $Sb_2Te_3$ nanostructures can be designed to exhibit interband plasmonic resonance. Figure 2b presents the experimental absorptance results for a visible metasurface design, showing 97% absorptance at 532 nm.

We further simulated the temperature distribution of thermoelectric detectors with and without the $Sb_2Te_3$ metasurface at an input heating power of 0.1 mW. As shown in Figure 2c, with the $Sb_2Te_3$ metasurface, the temperature difference in the detector device becomes more pronounced, reaching a maximum of 6.0 Kelvin due to the enhanced electric field in the metasurface region. In contrast, without the metasurface (Figure S5), the temperature distribution in $Sb_2Te_3$ is relatively uniform, with a temperature difference of 1.0 Kelvin. Figure 2d and S6 display the simulated potential distribution of these thermoelectric detectors under the same input power. Without the metasurface, the potential distribution is relatively flat, with the highest potential at the left end being 0.20 mV. With the $Sb_2Te_3$ metasurface, the potential gradient is enhanced, with the lowest potential at the metasurface reaching 0.40 mV, as holes migrate towards the cold end on the right. We could also note that improved responsivity is expected if a very small gold electrode is placed precisely at the hottest region of the metasurface, with another electrode positioned further away at room temperature. This analysis indicates that the $Sb_2Te_3$ metasurface enhances both the temperature gradient and the potential gradient, thereby improving the photothermoelectric conversion efficiency and providing an effective approach for optimizing the performance of photodetectors.

To demonstrate the enhanced PTE effect on metasurface photodetectors, we conducted the photovoltage tests on two types of detectors, with and without metasurfaces. As shown in Figure 2e, when the 532 nm laser is turned on, a voltage step of 2.4 mV is observed in the device with a metasurface; in comparison, when the laser illuminates the device with the flat $Sb_2Te_3$ film, a voltage step of only 0.65 mV is detected. In other words, the corresponding responsivities are 24 V/W and 6.5 V/W respectively, indicating a 4-fold increase in responsivity. When the laser irradiates from the metasurface region to the flat region, the photovoltage changes from positive to negative, as shown in Figure 2f. Under 532 nm laser illumination at the metasurface, the voltage reaches 2.2 mV. As the focused laser spot moves toward the flat region, the voltage gradually shifts, ultimately reaching 0.5 mV at the flat region. This measurement indicates a 4-fold voltage enhancement effect at the metasurface due to the resonance absorption, where $Sb_2Te_3$ nanostructures play a crucial role in thermoelectric detection, enhancing the thermoelectric response through interband plasmonic resonance.

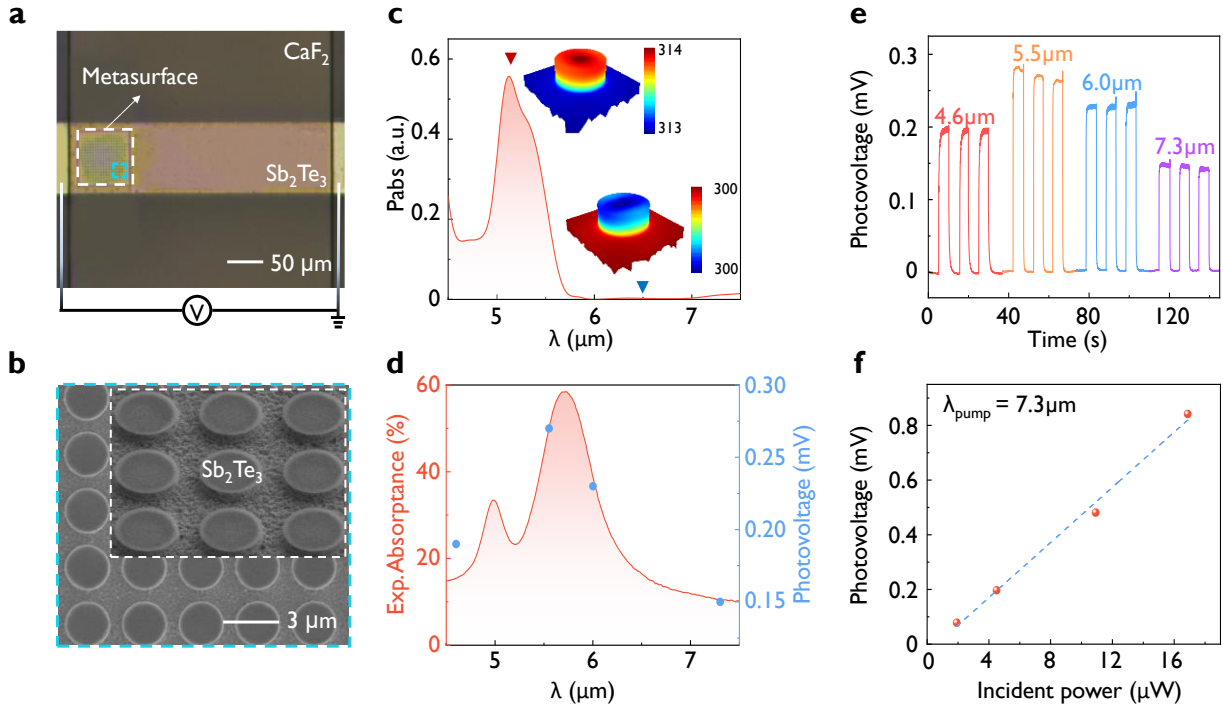

**Figure 3. Sb$_2$Te$_3$ mid-IR photodetector with hybridized resonance due to Mie and Fabry-Pérot (FP) cavities.** (a) Optical microscope image of the Sb$_2$Te$_3$ photodetector. The square region on the left end is the patterned metasurface regions. (b) SEM image of the fabricated mid-IR Sb$_2$Te$_3$ metasurface. (c) Simulated absorption power density with an incident power of 0.1 mW for the mid-IR metasurface. Insets show the temperature profile of the metasurface under corresponding wavelength conditions. (d) Measured absorptance spectrum of the mid-IR Sb$_2$Te$_3$ photodetector (disk diameter 2.0 µm, height 0.9 µm, pitch 3.0 µm, base Sb$_2$Te$_3$ film thickness of 150 nm). (e) Measured photovoltages with an incident laser power of 4.5 µW at different wavelengths (4.6, 5.5, 6.0, 7.3 µm). (f) Power-dependent photovoltage under illumination of incident optical powers from 1.9 to 16.8 µW.

Figure S7 shows the fabricated Sb$_2$Te$_3$ metasurface detector on a CaF$_2$ substrate, where it is connected to a printed circuit board (PCB) for measurement. By designing Sb$_2$Te$_3$ metasurfaces with different geometrical dimensions, multiple thermoelectric detectors can be fabricated on a single PCB board. Figure 3a presents the optical microscope image of showing mid-IR Sb$_2$Te$_3$ detector, which consists of the narrow strip of a width of 100 μm and a length of 400 μm. The enhanced absorption at the metasurface region increases the temperature gradient to result in a higher voltage. In addition, for the purpose of hot carrier transportation and the subsequent formation of voltage across the two electrodes, the mid-IR Sb$_2$Te$_3$ metasurface was formed by partial etching to a depth of 750 nm, leaving a 150-nm-thick base layer. Figure 3b shows the SEM images of the mid-IR Sb$_2$Te$_3$ nanostructures with a base layer.

Figures 3c show the simulated absorption power density of the Sb$_2$Te$_3$ metasurface in the mid-IR region, with the strongest absorption observed at 5.3 μm. The insets illustrate the simulated temperature profiles at 5.3 μm and 6.5 μm, where the absorption power density at 5.3 μm is higher, leading to a temperature gradient of up to 0.35 Kelvin/μm, indicating the effective conversion of incident light into heat by the metasurface. Figure 3d presents the measured absorptance of the sample, reaching 60% at 5.9 μm. By varying the diameter of the Sb$_2$Te$_3$ structures from 1.4 μm to 2.0 μm, there exists a dual absorption modulation, ranging from 4.3 to 5.9 μm (see details in Figure S8). To evaluate the wavelength-selective detection capability and high sensitivity at room temperature, a series of photoelectric tests were performed using a series of quantum cascade lasers (QCL, Daylight Solutions, MIRcat) of different wavelengths and intensities. For instance, Figure 3e shows the response of the Sb$_2$Te$_3$ metasurface device to the laser wavelength of 4.6, 5.5, 6.0, and 7.3 μm, with a constant power of 4.5 μW. Notably, the voltage response to the 5.5 μm laser is the highest, reaching 300 μV, and the responsivity reached to 67 V/W, due to the peak absorption of the device at this wavelength. In other words, our $Sb_2Te_3$ detector can achieve wavelength-sensitive detection in the mid-IR range. Additionally, the impact of incident power level on the photovoltage was analyzed, as shown in Figure 3f. The photovoltage varies with increasing laser power. Through linear fitting, we obtained $R^2 = 0.997$ and the $V_{PTE} = 0.05 P_{Incident} - 0.031$, indicating a linear relationship between the photovoltage and incident light power. This trend highlights the amplification of the photoelectric effect with increased laser power, leading to increased absorption by the $Sb_2Te_3$ nanostructures, thereby increasing the thermal gradient and consequently the photovoltage.

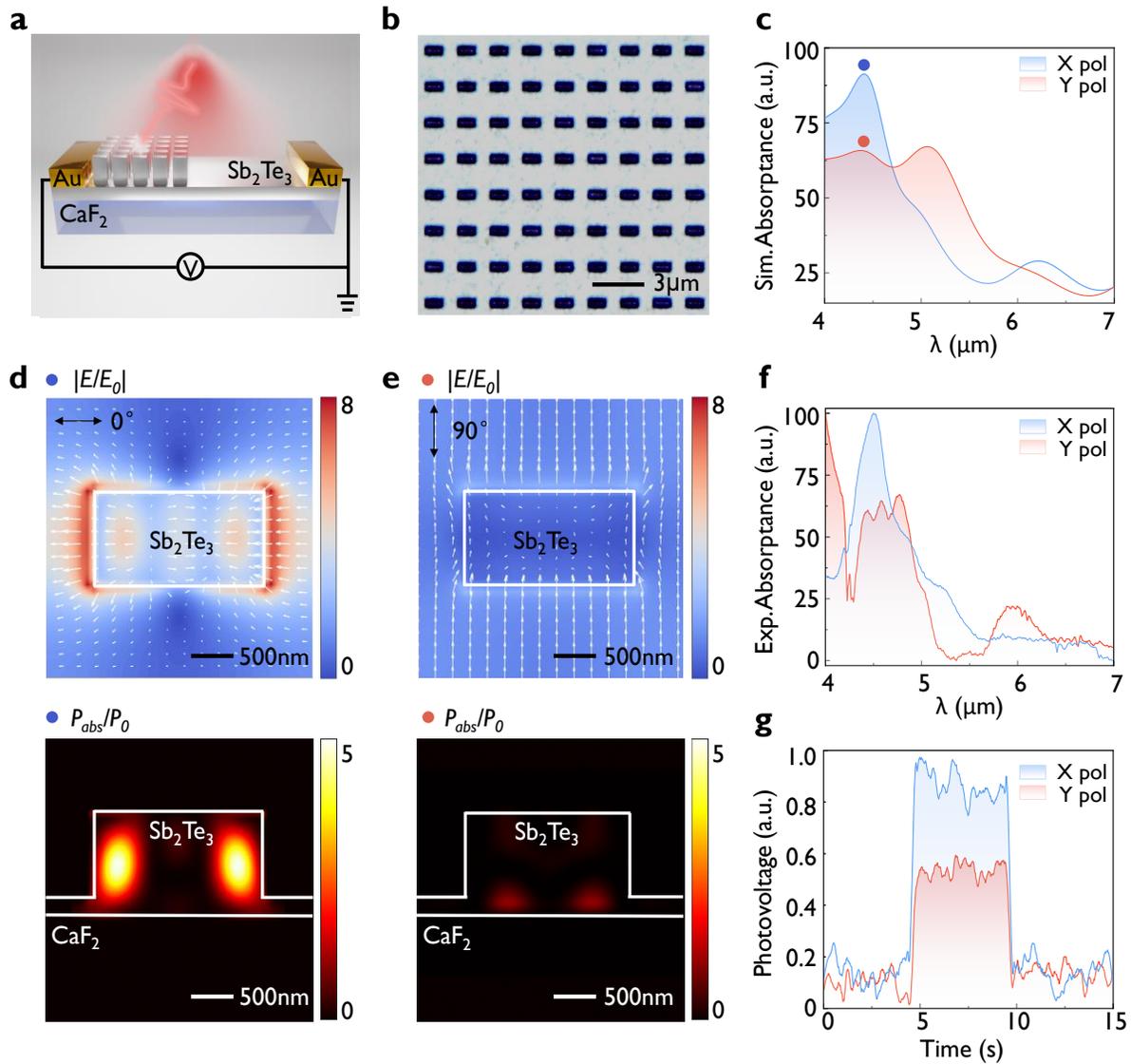

**Figure 4. Polarization-sensitive Sb₂Te₃ mid-IR photodetector.** (a) Schematic of polarized Sb$_2$Te$_3$ infrared photodetector. (b) Optical microscope image of the polarization dependent Sb$_2$Te$_3$ metasurface. (c) Simulated polarization-dependent absorptance. (d-e) Simulated electric field distributions ($|E/E_0|$) and absorption intensity distribution($|P/P_0|$) for *x*- and *y*-polarizations. (f) Measured polarization-dependent absorptance by Fourier Transform Infrared Spectroscopy (FTIR). (g) Photovoltage measurements when the laser shines on the nanostructure under 4.5 μm laser with different polarization.

Figure 4a illustrates the schematic diagram of a polarization-sensitive $Sb_2Te_3$ mid-IR photodetector. In addition, Figure 4b presents an optical image of the array of fabricated $Sb_2Te_3$ polarization-dependent metasurfaces, clearly displaying the well-defined morphology of 2.0-µm-long and 1.0-µm-wide $Sb_2Te_3$ rectangular arrays with a period of 3.0 µm. Here, the base layer of $Sb_2Te_3$ film is also 150 nm. Figure 4c and Figure 4f presented the measured and simulated absorption spectra, achieving an absorption difference at 4.5 µm. In addition, the simulated electric field distribution for different polarizations is shown in Figure 4d and 4e, where the electric field amplitude and absorption intensity distribution at 0° polarization is higher than the one under 90° polarization.

The electrical response of detector to 4.5 µm polarized light is shown in Figure 4g. The photovoltage measurements of the nanostructures indicate that the response at 0° polarization is 8.0% higher than at 90° polarization, which is close to the 40% absorption difference measured at the same wavelength for different polarizations. To verify the polarization sensitivity of the metasurface to different laser wavelengths, lasers with wavelengths of 4.5, 4.8, 5.5, and 6.0 µm were used to irradiate the sample and measure the corresponding photovoltage. As shown in Figure S9, the 4.5 µm laser exhibited the highest polarization sensitivity and absorption difference, corresponding to the metasurface absorption measured in the experiment. This measurement demonstrates that by designing different $Sb_2Te_3$ metasurfaces, wavelength and polarization-selective absorption in the mid-IR region at room temperature can be achieved.

**Conclusion**

We demonstrate that the chalcogenide antimony telluride ($Sb_2Te_3$) is a multi-functional material platform for photodetection applications, due to its excellent thermoelectric property as well as its rich optical characteristics across multiple spectral ranges, from visible to mid-IR. Through the systematic designs, we have demonstrated the nanostructured $Sb_2Te_3$ metasurface detectors, working in both visible and mid-IR ranges, where the polarization-sensitive characteristic can also be enabled by designing the asymmetric metasurface. This work advances the integration of thermoelectric and optical functionalities of $Sb_2Te_3$ through nanofabrication, providing a novel platform for single-material ultra-broadband metasurfaces and offering new directions for enable the miniaturized on-chip detection of the higher-dimensional characteristics of light across an ultra-wideband spectrum wavelength range.

**METHODS**

**$Sb_2Te_3$ Sputtering *via* Radio Frequency (RF) Sputtering.** A $Sb_2Te_3$ film (1µm) was deposited on $CaF_2$ substrate (10mm × 10mm × 0.5mm) by using radio frequency (RF) sputtering. The film was deposited from a 2-inch diameter $Sb_2Te_3$ target (purity > 99.9%) with a power of 30 W and a pressure of 3.7 mTorr pure Ar at room temperature. The deposition rate is 4.725 nm/minute.

**Nanopatterning of $Sb_2Te_3$ Metasurfaces.** A ~30 nm Hydrogen silsesquioxane (HSQ) mask was created on the surface of $Sb_2Te_3/CaF_2$ sample by using an electron beam lithography (EBL, Elionix ELS-7000). First, HSQ resist (Dow Corning XR-1541-006) was spin-coated onto a cleaned sample at 3000 round-per-minute (rpm) to obtain a HSQ thickness of ~100 nm. Then, electron beam exposure was carried out with an electron acceleration voltage of 100 keV and a beam current of 500 pA. After completing the exposure, the pattern was immediately developed by NaOH/NaCl salty solution (1% wt./4% wt. in de-ionized water) for one minute. After that, the sample was flushed by de-ionized water for one minute to stop the development. Finally, the sample was rinsed by using isopropanol alcohol (IPA) and dried by a continuous flow of nitrogen gas. $Sb_2Te_3$ etching was then carried out by inductively-coupled-plasma (ICP, Oxford Instruments Plasmalab System 100), with a RF power of 100 watts, ICP power of 500 watts, HBr with a flow rate of 50 sccm (standard-cubic-centimeters-per-minute), Ar with a flow rate of 5 sccm under a process pressure of 5 mTorr, and a temperature of 50 °C.

**Photolithography, Dry Etching and Metal Contact Fabrication.** The metasurface sample was coated with hexamethyldisilizane (HMDS) to enhance the adhesion of the photoresist with the sample surface. Then the sample was spin coated with photoresist, S1811 at 3000 rpm. After spin coating, the sample undergoes a soft-baking process for 1 minute at 100°C. This results in a photoresist layer with thickness of ~1.5 µm. After that, the sample was aligned under the desired pattern drawn on a chromium mask, using the mask aligner system (EVG6200). Ultraviolet light with a dose of 50 mJ/cm² was exposed onto the photoresist. Finally, the sample is developed with MF319 for 1.5 minutes to create the pattern. $Sb_2Te_3$ strip was etched by using inductively-coupled-plasma (ICP, Oxford Instruments Plasmalab System 100) with the same condition in dry Etching of $Sb_2Te_3$ metasurface. Then the remaining photoresist is cleaned by using acetone. The procedure of photolithography is the same as mask creation of $Sb_2Te_3$ strip. Here, the chromium mask for electrode contact is used. 20-nm-thick Ti was deposited by electron beam evaporation (Denton Explorer), at a deposition rate of 1 angstrom per second, and at a pressure of $9\times10^{-7}$ Torr. Then, 60 nm Au was deposited using the same system, at a deposition rate of 1 angstrom per second, and at a pressure of $9\times10^{-7}$ Torr.

**Characterizations.** The experimental mid-IR spectra were obtained using a microscope-based Fourier-transform infrared (FTIR, Bruker Hyperion 2000) spectroscopy system, operating in reflection mode. The system employed a globar as the illumination source, providing a stable and broad-spectrum infrared light necessary for accurate spectral analysis. The detection of the reflected infrared light was carried out using a mercury cadmium telluride (MCT) detector, which was cooled with liquid nitrogen to enhance its sensitivity and to reduce its thermal noise, thereby improving the accuracy and reliability of the measurements. To investigate the polarization properties of the samples, a wire-grid polarizer was incorporated into the experimental setup. The visible reflectance spectra of $Sb_2Te_3$ nanodisk arrays were measured using a Craic micro-spectrometer, with a ×5 objective lens with a numerical aperture of 0.12. Moreover, the optical images of the color palettes were captured by an Olympus microscope (MX61) using the "analySIS" software. The objective was a ×10 MPlanFLN, NA=0.30 lens. Before the image was captured, a white color balancing was carried out on a 100-nm-thick aluminum film, which was prepared by electron beam evaporation. SEM images were taken at an acceleration voltage of 10 keV with the Elionix, ESM-9000 SEM.

The visible photoresponse is measured by using WITec Alpha 300 S Optical Microscope in confocal mode. A 532 nm CW laser was focused on the sample. The generated photovoltage was then recorded by sourcemeter (Keithley 2636B). The mid-IR photoresponse is measured by using a homemade photocurrent measurement system where the infrared light with different polarization status is obtained from a serious quantum cascade lasers (Daylight Solutions, MIRcat) and tunable wavelength in the range of 4-8 μm combining a serious half-wave plate and quarter-wave plate, and then focused on the samples using a zinc selenide IR focusing lens with a focal length of 50 mm. The generated photovoltage was then recorded by a highly sensitive source meter unit (Keysight, B2912A).

**Numerical Simulations.** The optical simulations and power absorption density of the $Sb_2Te_3$ nanodisk arrays were performed using Lumerical finite-difference time-domain (FDTD) solver. A plane source with wavelengths between 380 nm and 11 μm was input in the negative z-direction perpendicular to the nanodisk array and substrate. A field monitor was placed 100 nm above the plane source to measure the reflected power, while another field monitor was placed perpendicular to the substrate along *x*=0 nm to measure the electric and magnetic fields in the cross-section of the nanodisk. Periodic boundary conditions were set for the *x*- and *y*-boundaries, while the perfectly matched layer (PML) boundary condition was set for the *z*-boundaries. Refractive index data for the polycrystalline phase of $Sb_2Te_3$ were measured using an ellipsometer as shown in Figure 1(b). Power absorption density is calculated by $P_{abs} = 1/2\omega\varepsilon''|E|^2$, and is normalized by $P_0$, the incident power divided by the meta-molecule volume. Multipole decomposition analysis and thermal simulation were carried out using the COMSOL Multiphysics 5.6 Optical Wave Optics Module and heat transfer modules, which exploits the Finite Element Method. Because of differences in length scales, first electromagnetic field simulations were computed with smaller simulation regions, and absorbed power was input into a heat transfer simulation that had a much larger simulation area. A single unit-cell of the structure was simulated using periodic boundary conditions under normally incident plane-wave.

**ASSOCIATED CONTENTS**

**Supporting Information**

Figure S1: Characterization of $Sb_2Te_3$ film.

Figure S2: The refractive index (n) and extinction coefficient (k) of $Sb_2Te_3$ in visible region.

Figure S3: Multi-physical decomposition of the $Sb_2Te_3$ metasurface interactions.

Figure S4: The refractive index (n) and extinction coefficient (k) of $Sb_2Te_3$ in near-IR range.

Figure S5: Simulated temperature gradient for the thermoelectric detector without the $Sb_2Te_3$ metasurface.

Figure S6: Simulated potential gradient for the thermoelectric detector without the $Sb_2Te_3$ metasurface.

Figure S7: Photograph of the actual $Sb_2Te_3$ detector device.

Figure S8: Measured reflectance of $Sb_2Te_3$ mid-IR metasurface with various diameters.

Figure S9: The polarization standard deviation of the photovoltage measurement at different wavelengths.

**AUTHOR INFORMATION**

**Corresponding Author**

*E-mail: dongz@imre.a-star.edu.sg.

*E-mail: joel_yang@sutd.edu.sg.

*E-mail: qjwang@ntu.edu.sg.

*E-mail: r.e.simpson.1@bham.ac.uk.

**ORCID**

Shutao Zhang: 0000-0002-9343-5550

Joel K. W. Yang: 0000-0003-3301-1040

Zhaogang Dong: 0000-0002-0929-7723

**Author contributions**

J.K.W.Y. and Z.D. conceived the concepts and supervised the project. R.E.S. conceived the concept of mid-IR $Sb_2Te_3$ detector. S.Z., S.A., J.Z., H.L.Y.L. and Z.D. prepared the samples. S.Z. and J.Z. did the SEM characterizations and interpretations. S.A. and S.Z. did the electron beam lithography, dry etching and developed the whole fabrication processes. M.D. and Q.J.W. did the mid-IR device characterizations and interpretations. S.Z. and Y.L. did the numerical simulations. N.Q.A. did the sputtering of $Sb_2Te_3$ films, under the guidance of R.E.S. Y.L. and S.Z. did thermal simulation and analysis. S.Z. and Q.Y.S.W. did the optical reflectance measurements. S.A., Y.L. and Z.D. provided expertise in data analysis and interpretations. N.L.M.W. did the mid-IR ellipsometer measurement and interpretation. A.S. did the characterization of the Seebeck coefficient and interpretation. J.D. participated in the discussions and provided the suggestions. The paper was drafted by S.Z with input from Z.D., J.K.W.Y. and S.A. All authors analyzed the data and read and corrected the manuscript before the submission. S.Z. and S.A. are equal contributions.

**Notes**

The authors declare no competing financial interests.

**ACKNOWLEDGMENTS**

Z.D. and J.K.W.Y. would like to acknowledge the funding support from A*STAR AME IRG (Project No. A20E5c0093) and National Research Foundation (NRF) via Grant No. NRF-CRP30-2023-0003. In addition, Z.D. would like to acknowledge the funding support from the Agency for Science, Technology and Research (A*STAR) under its Career Development Award grant (Project No. C210112019), MTC IRG (Project No. M21K2c0116 and M22K2c0088), the Quantum Engineering Program 2.0 (Award No. NRF2021-QEP2-03-P09) and DELTA-Q 2.0 (Project No. C230917005). J.K.W.Y. would like to acknowledge the funding support from National Research Funding (NRF) Singapore NRF-CRP20-2017-0001 and NRF-NRFI06-2020-0005. Q.Y.S.W would like to acknowledge funding support from The National Research Foundation, Singapore under NRF-CRP (NRF-CRP26-2021-0004).

*Supplementary Information for*

# Chalcogenide Metasurfaces Enabling Ultra-Wideband Detectors from Visible to Mid-infrared

Shutao Zhang[1,2,3,#], Shu An[1,#], Mingjin Dai[4], Qing Yang Steve Wu[1], Nur Qalishah Adanan[2], Jun Zhang[1], Yan Liu[1], Henry Yit Loong Lee[1], Nancy Lai Mun Wong[1], Ady Suwardi[1,5], Jun Ding[3], Robert Edward Simpson[2,6,*], Qi Jie Wang[4,*], Joel K. W. Yang[2,*] and Zhaogang Dong[1,2,3,*]

[1]Institute of Materials Research and Engineering (IMRE), Agency for Science, Technology and Research (A*STAR), 2 Fusionopolis Way, Innovis #08-03, Singapore 138634, Republic of Singapore

[2]Singapore University of Technology and Design (SUTD), 8 Somapah Road, 487372, Singapore

[3]Department of Materials Science and Engineering, National University of Singapore, 9 Engineering Drive 1, Singapore 117575

[4]School of Electrical and Electronic Engineering, Nanyang Technological University, Singapore 639798, Singapore

[5]Department of Electronic Engineering, The Chinese University of Hong Kong, Sha Tin, New Territories, Hong Kong SAR 999077, China

[6]University of Birmingham, Edgbaston, B15 2TT, UK

[#]These authors equally contribute to this work.

*Correspondence and requests for materials should be addressed to J.K.W.Y. (email: joel_yang@sutd.edu.sg), Q.J.W. (email: qjwang@ntu.edu.sg), R.E.S. (email: r.e.simpson.1@bham.ac.uk) and Z.D. (email: dongz@imre.a-star.edu.sg).

**Figure S1**: Characterization of $Sb_2Te_3$ film.

**Figure S2**: The refractive index (n) and extinction coefficient (k) of $Sb_2Te_3$ in visible region.

**Figure S3**: Multi-physical decomposition of the $Sb_2Te_3$ metasurface interactions.

**Figure S4**: The refractive index (n) and extinction coefficient (k) of $Sb_2Te_3$ in near-IR range.

**Figure S5**: Simulated temperature gradient for the thermoelectric detector without the $Sb_2Te_3$ metasurface.

**Figure S6**: Simulated potential gradient for the thermoelectric detector without the $Sb_2Te_3$ metasurface.

**Figure S7**: Photograph of the actual $Sb_2Te_3$ detector device.

**Figure S8**: Measured reflectance of $Sb_2Te_3$ mid-IR metasurface with various diameters.

**Figure S9**: The polarization standard deviation of the photovoltage measurement at different wavelengths.

## S1. Characterization of Sb$_2$Te$_3$ film.

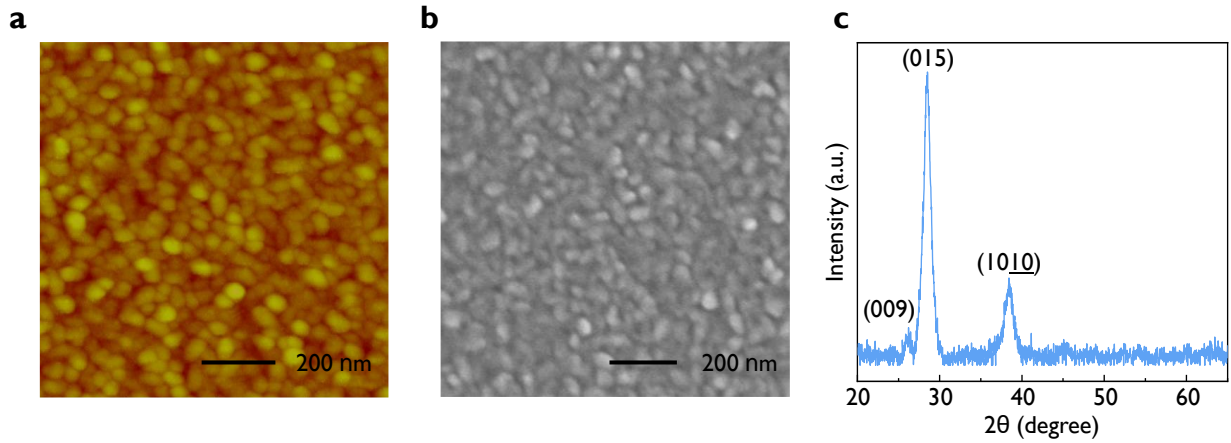

Fig. S1. (a) AFM images of the Sb$_2$Te$_3$ as-deposited film, showing a surface roughness of 3.2 nm. (b) SEM images of the as-deposited film. (c) XRD characterization of the as-deposited film, indicating that the layered structure of Sb$_2$Te$_3$ exhibits a pronounced (015) orientation.

**S2. The refractive index (n) and extinction coefficient (k) of Sb$_2$Te$_3$ in visible region.**

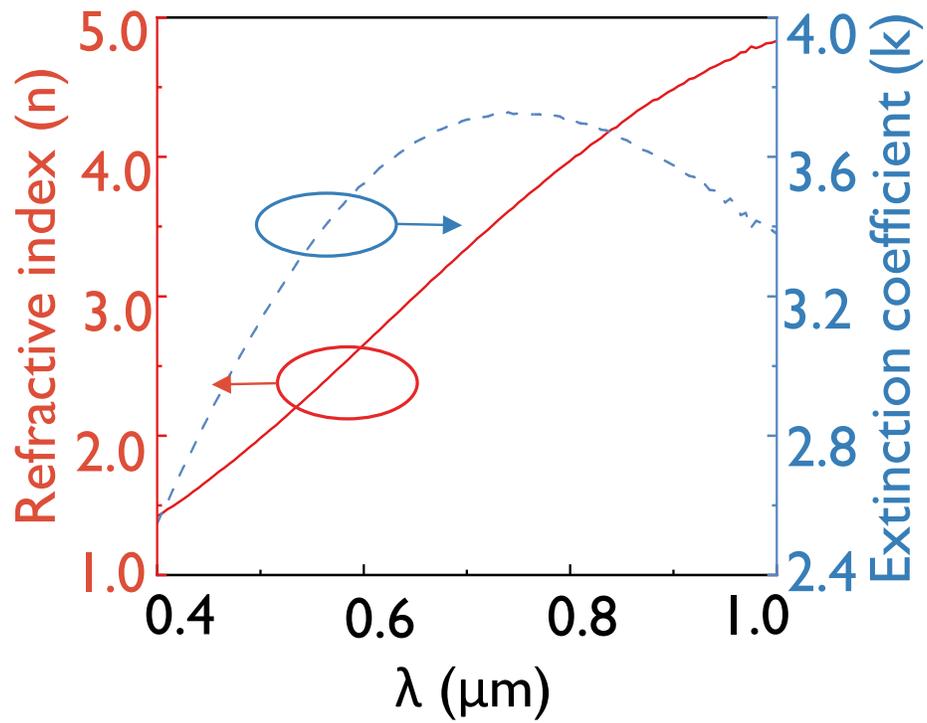

Fig. S2. The refractive index (n) and extinction coefficient (k) of Sb$_2$Te$_3$ in visible region.

**S3. Multi-physical decomposition of the Sb₂Te₃ metasurface interactions.**

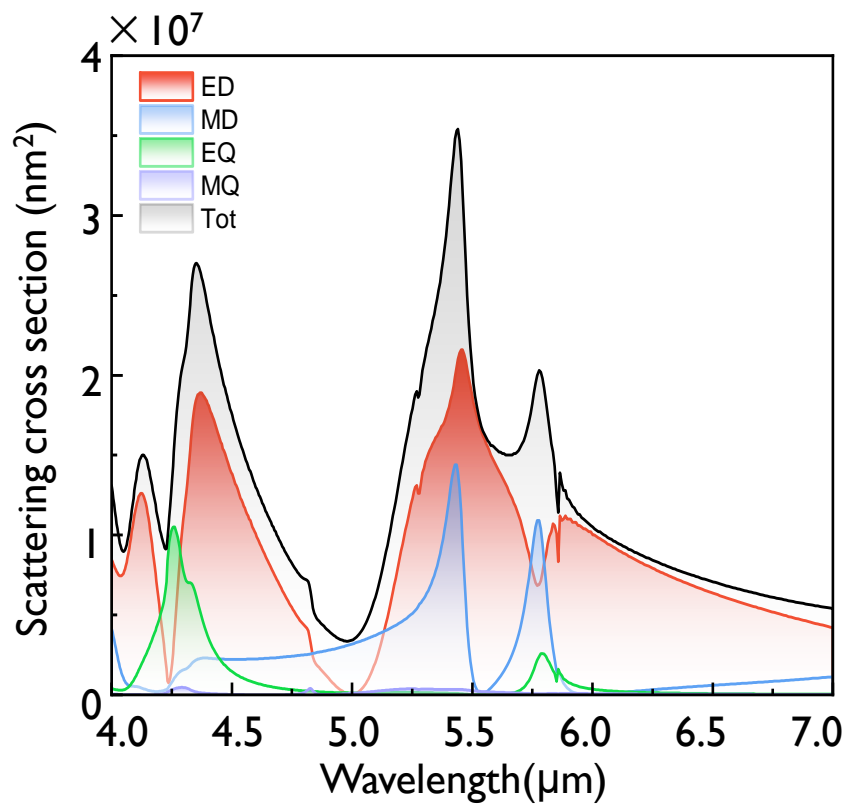

Fig. S3. Multi-physical decomposition of the metasurface interactions computed using COMSOL Multiphysics, showing contributions of electric dipole (ED), magnetic dipole (MD), electric quadrupole (EQ), magnetic quadrupole (MQ), and the total combined effect (Tot).

**S4: The refractive index (n) and extinction coefficient (k) of Sb₂Te₃ in near-IR range.**

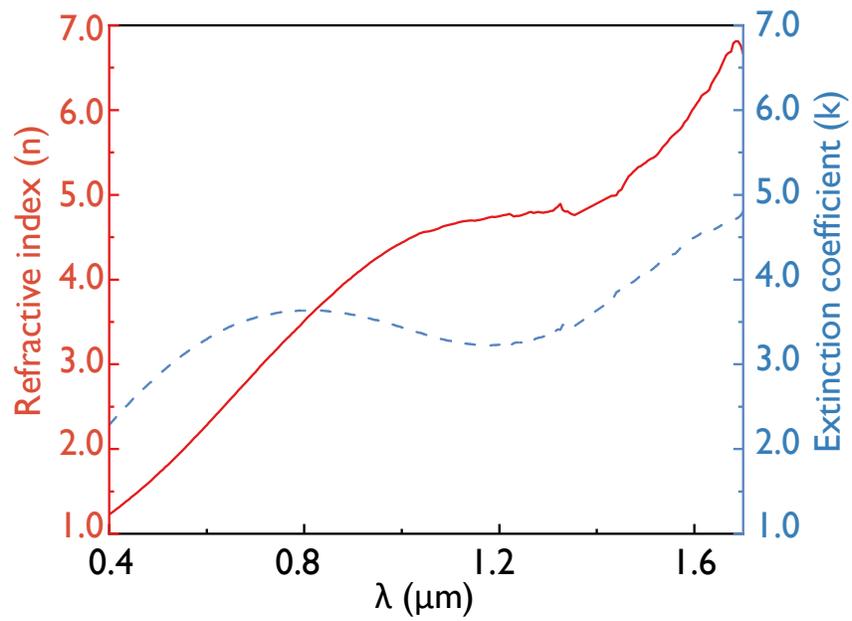

Fig. S4. The refractive index (*n*) and extinction coefficient (*k*) of Sb₂Te₃ in near-IR range.

**S5. Simulated temperature gradient for the thermoelectric detector without the Sb$_2$Te$_3$ metasurface.**

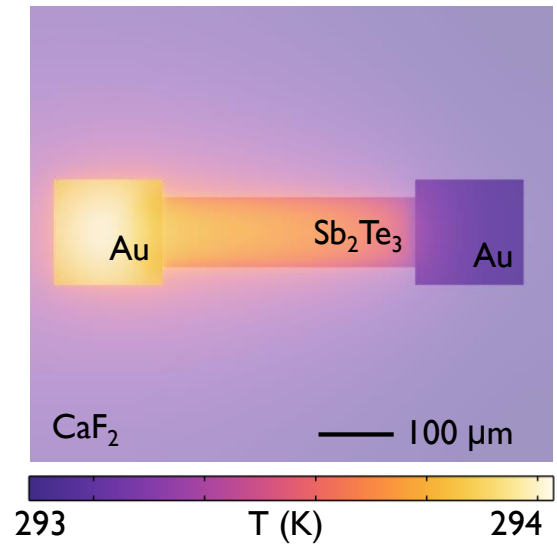

Fig. S5. Simulated temperature gradient with an input power of 0.1 mW. Temperature distribution for the thermoelectric detector without the Sb$_2$Te$_3$ metasurface.

## S6. Simulated potential gradient for the thermoelectric detector without the $Sb_2Te_3$ metasurface.

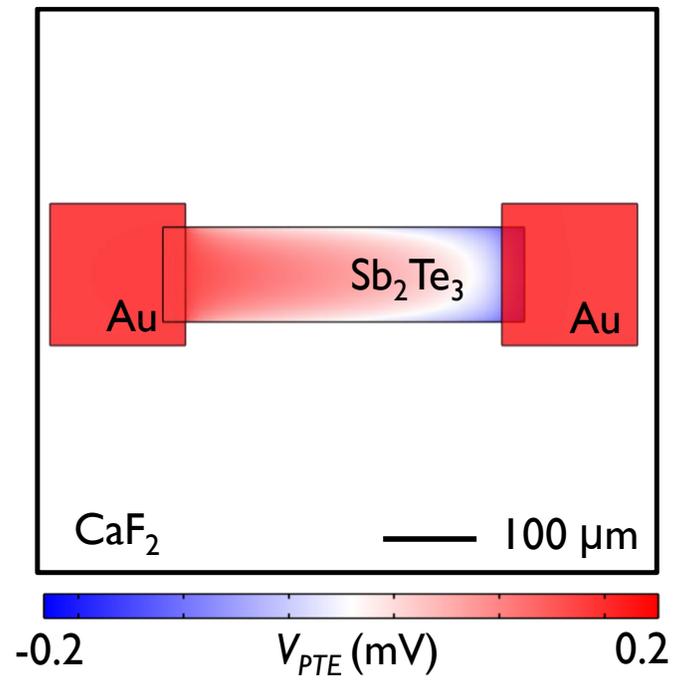

Fig. S6. Simulated potential with an input power of 0.1 mW. Potential gradient for the thermoelectric detector without the $Sb_2Te_3$ metasurface.

**S7. Photograph of the actual Sb₂Te₃ detector device.**

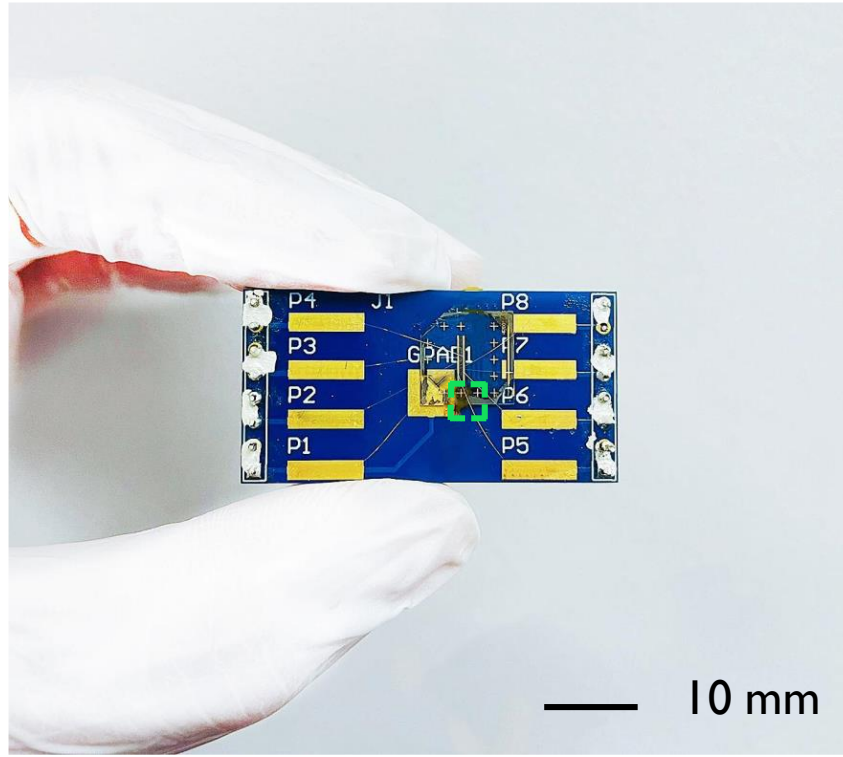

Fig. S7. Photograph of the actual mid-IR detector devices wirebonded to pads on a printed circuit board (PCB), and an optical microscope image of the Sb$_2$Te$_3$ photodetector in the center panel.

**S8. Measured reflectance of Sb₂Te₃ mid-IR metasurface with various diameters.**

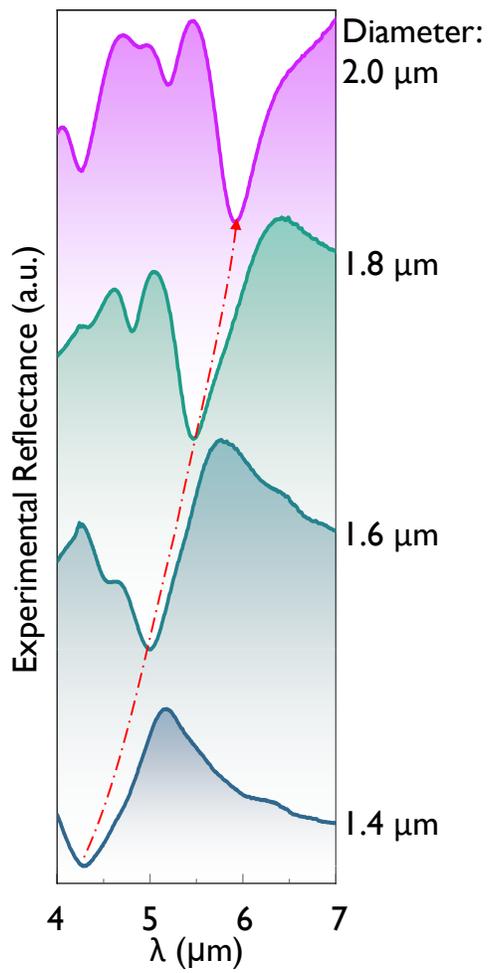

Fig. S8. Measured reflectance of nanodisks with various diameters (Diameter = 1.4-2.0 μm, step size = 0.2 μm).

**S9. The polarization standard deviation of the photovoltage measurement at different wavelengths.**

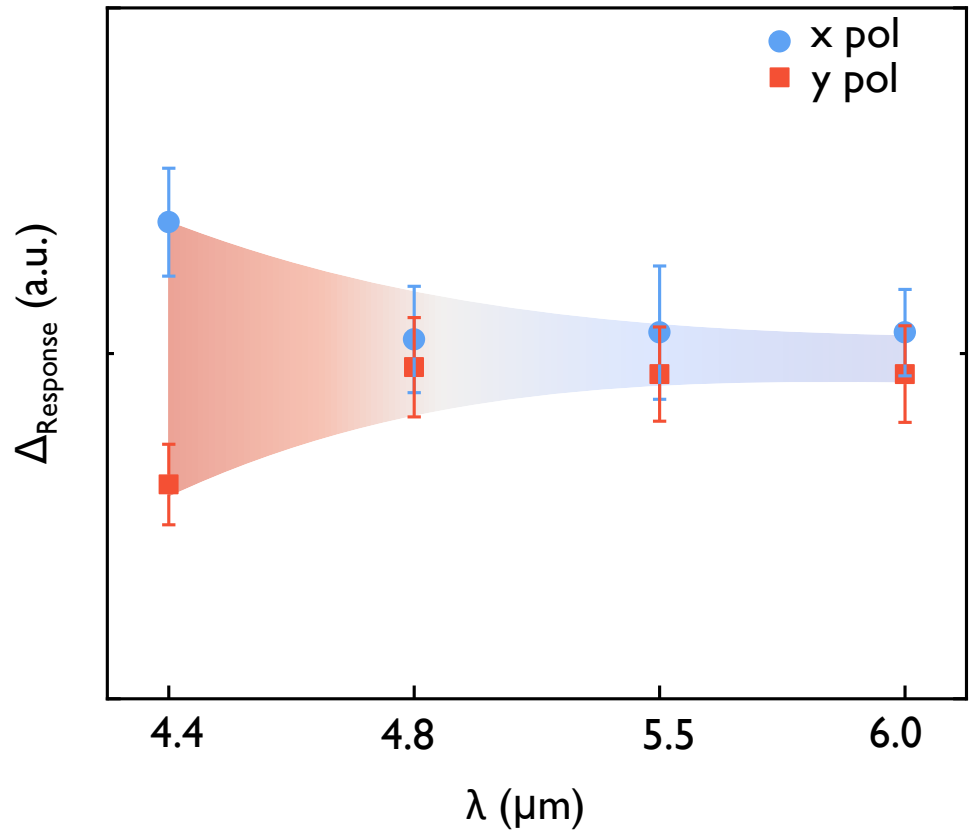

Fig. S9. The polarization standard deviation of the photovoltage measurement when the polarization-sensitive $Sb_2Te_3$ mid-IR photodetector is illuminated by laser working at different wavelengths.